\documentclass[runningheads]{llncs}
\usepackage{amsmath,graphicx}
\usepackage{amssymb}
\usepackage{xcolor}
\usepackage{diagbox}
\usepackage{booktabs} 
\usepackage{listings}
\usepackage{xcolor}

\definecolor{codegreen}{rgb}{0,0.6,0}
\definecolor{codegray}{rgb}{0.5,0.5,0.5}
\definecolor{codepurple}{rgb}{0.58,0,0.82}
\definecolor{backcolour}{rgb}{0.95,0.95,0.92}

\lstdefinestyle{mystyle}{
    backgroundcolor=\color{backcolour},   
    commentstyle=\color{codegreen},
    keywordstyle=\color{magenta},
    numberstyle=\tiny\color{codegray},
    stringstyle=\color{codepurple},
    basicstyle=\ttfamily\scriptsize,
    breakatwhitespace=false,         
    breaklines=true,                 
    captionpos=b,                    
    keepspaces=true,                 
    numbers=left,                    
    numbersep=5pt,                  
    showspaces=false,                
    showstringspaces=false,
    showtabs=false,                  
    tabsize=2
}

\begin{document}
\title{LambdaUNet: 2.5D Stroke Lesion Segmentation of Diffusion-weighted MR Images }
\titlerunning{LambdaUNet: 2.5D Stroke Lesion Segmentation of DWIs}
\author{Anonymous Authors}
\author{Yanglan Ou\inst{1}, Ye Yuan\inst{2}, Xiaolei Huang\inst{1}, Kelvin Wong\inst{3}, \break John Volpi\inst{4}, James Z. Wang\inst{1}, Stephen T.C. Wong\inst{3}}
\institute{The Pennsylvania State University, University Park, Pennsylvania, USA \and
Carnegie Mellon University, Pittsburgh, Pennsylvania, USA \and
TT and WF Chao Center for BRAIN \& Houston Methodist Cancer Center, Houston Methodist Hospital, Houston, Texas, USA \and
Eddy Scurlock Comprehensive Stroke Center, Department of Neurology,\break Houston Methodist Hospital, Houston, Texas, USA}
\authorrunning{Y. Ou et al.}

\maketitle              %
\begin{abstract}
Diffusion-weighted (DW) magnetic resonance imaging is essential for the diagnosis and treatment of ischemic stroke. DW images (DWIs) are usually acquired in multi-slice settings where lesion areas in two consecutive 2D slices are highly discontinuous due to large slice thickness and sometimes even slice gaps. Therefore, although DWIs contain rich 3D information, they cannot be treated as regular 3D or 2D images. Instead, DWIs are somewhere in-between (or 2.5D) due to the volumetric nature but inter-slice discontinuities. Thus, it is not ideal to apply most existing segmentation methods as they are designed for either 2D or 3D images. To tackle this problem, we propose a new neural network architecture tailored for segmenting highly-discontinuous 2.5D data such as DWIs. Our network, termed \texttt{LambdaUNet}, extends UNet by replacing convolutional layers with our proposed \texttt{Lambda+} layers. In particular, \texttt{Lambda+} layers transform both intra-slice and inter-slice context around a pixel into linear functions, called lambdas, which are then applied to the pixel to produce informative 2.5D features. \texttt{LambdaUNet} is simple yet effective in combining sparse inter-slice information from adjacent slices while also capturing dense contextual features within a single slice. Experiments on a unique clinical dataset demonstrate that {\texttt{LambdaUNet}} outperforms existing 3D/2D image segmentation methods including recent variants of UNet. Code for \texttt{LambdaUNet} is available.\footnote{URL: \url{https://github.com/YanglanOu/LambdaUNet}}

\end{abstract}
\begin{figure}
\vspace{-5mm}
\includegraphics[width=\textwidth]{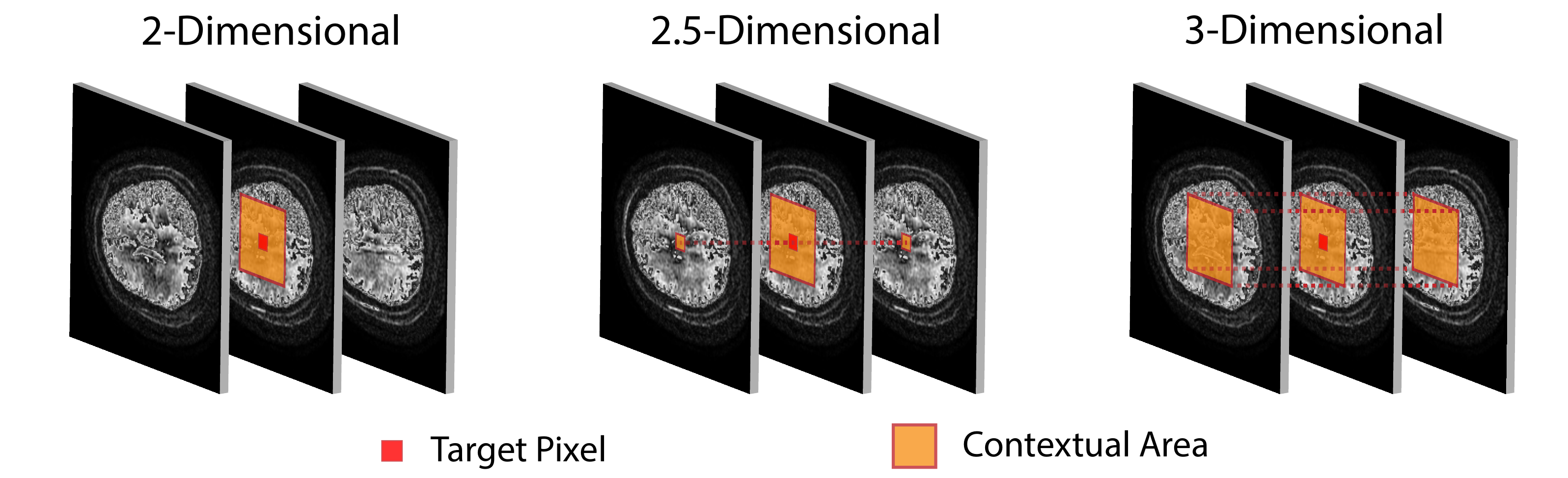}
\vspace{-7mm}
\caption{Comparison of 2D, 2.5D, and 3D feature extraction methods. When extracting features for a target pixel, our 2.5D method restricts the context area in adjacent slices to focus on the most relevant pixels to reduce noise and improve generalization.} \label{fig:overview}
\vspace{-5mm}
\end{figure}

\section{Introduction}
\label{sec:intro}
\vspace{-1mm}
In the United States, stroke is the second leading cause of death and the third leading cause of disability~\cite{johnson2016stroke}. About 795,000 people in the US have a stroke each year~\cite{mozaffarian2016heart}. A stroke happens when some brain cells suddenly die or are damaged due to lack of oxygen when blood flow to parts of the brain is lost or reduced due to blockage or rupture of an artery~\cite{owolabi2015burden}. Locating the lesion areas where brain tissue is prevented from getting oxygen and nutrients is essential for accurate evaluation and timely treatment. Diffusion-weighted imaging (DWI) is a commonly performed magnetic resonance imaging (MRI)  
sequence for evaluating acute ischemic stroke and is sensitive in detecting small and early infarcts~\cite{lansberg2002diffusion}. 

Segmenting stroke lesions on DWIs manually is time-consuming and subjective~\cite{kanchana2020ischemic}. With the advancement of deep learning, numerous automatic segmentation methods based on deep neural networks (DNNs) have emerged to detect stroke lesions. Some of them perform segmentation on each 2D slice individually~\cite{charoensuk2015acute,chen2017fully}, while others treat DWIs as 3D data and apply 3D segmentation networks~\cite{zhang2018automatic}. Beyond methods for lesion segmentation in DWIs, there have been many successful methods for general medical image segmentation. For instance, UNet~\cite{ronneberger2015u} has shown the advantage of skip-connections on biomedical image segmentation. Based on UNet, Oktay~\emph{et al.} proposed Attention UNet by adding attention gates that filter the features propagated through the skip connections in U-Net~\cite{oktay2018attention}; Chen~\emph{et al.} proposed TransUNet, as they find that transformers make strong encoders for medical image segmentation~\cite{chen2021transunet}. \text{Çiçek}~\cite{cciccek20163d} extend UNet to 3D field for volumetric segmentation. Wang~\emph{et al.} proposed volumetric attention combined with Mask-RCNN to address the GPU memory limitation of 3D U-net. Zhang~\emph{et al.}~\cite{zhang2018automatic} proposed a 3D fully-convolutional and densely-connected convolutional network which is derived from the powerful DenseNet~\cite{huang2017densely}.

\setcounter{footnote}{0} 
Although previous medical image segmentation methods work well for 2D or 3D data by design, they are not well suited for DWIs, which have contextual characteristics between 2D and 3D. We term such data type as 2.5D~\cite{zhang2019multiple}.\footnote{Note that our definition of 2.5D is different from that in computer vision, where 2.5D means the 2D retinal projections of 3D environments.} Different from 2D data, DWIs contain 3D volumetric information by having multiple DWI slices.
However, unlike typical 3D medical images that are isotropic or near isotropic in all three dimensions, DWIs are highly discontinuous between slices; neighboring slices can have abrupt changes around the same area, and 
lesions in early infarcts are often small and do not even extend beyond a few slices. Due to the 2.5D characteristics of DWIs, if we apply 2D segmentation methods to DWIs, we lose valuable 3D contextual information from neighboring slices (Fig.~\ref{fig:overview}\,(left)). On the other hand, if we apply a traditional 3D CNN-based segmentation method, due to the high discontinuity between slices, many irrelevant features from neighboring slices are processed by the network (Fig.~\ref{fig:overview}\,(right)), which adds substantial noise to the learning process and also makes the network prone to over-fitting.

In this work, 
our goal is to design a segmentation network tailored for images with 2.5D characteristics like DWIs. To this end, we propose \texttt{LambdaUNet} which adopts the UNet~\cite{ronneberger2015u} structure but replaces convolutional layers with our proposed \texttt{Lambda+} layers which can capture both dense intra-slice features and sparse inter-slice features effectively. \texttt{Lambda+} layers are inspired by the Lambda layers~\cite{bello2021lambdanetworks} which transform both global and local context around a pixel into linear functions, called lambdas, and produce features by applying these lambdas to the pixel. Although Lambda layers have shown strong performance for 2D image classification, they are not suitable for 2.5D DWIs because they are designed for 2D data and cannot capture sparse inter-slice features. Our proposed \texttt{Lambda+} layers are designed specifically for 2.5D DWI data, where they consider both the intra-slice and inter-slice contexts of each pixel. Here the inter-slice context of a pixel consists of pixels at the same 2D location but in neighboring slices (Fig.~\ref{fig:overview}\,(middle)). Note that, unlike many 3D feature extraction methods, %
\texttt{Lambda+} layers do not consider pixels in neighboring slices that are at different 2D locations, because these pixels are less likely to contain relevant features and we suppress them to reduce noise and prevent over-fitting. \texttt{Lambda+} layers transform the inter-slice context into a different linear function--inter-slice lambda--which complements other intra-slice Lambdas to derive sparse inter-slice features. As illustrated in Fig.~\ref{fig:overview}, the key design of \texttt{Lambda+} layers is that they treat intra-slice and inter-slice features differently by using a dense intra-slice context and a sparse inter-slice context, which suits well the 2.5D DWI data. 

To our knowledge, the proposed \texttt{LambdaUNet} is the first 2.5D segmentation model designed specifically for 2.5D data like DWIs.
Extensive experiments on a large annotated clinical DWI dataset of stroke patients show that \texttt{LambdaUNet} significantly outperforms previous art in terms of segmentation accuracy.

\vspace{-3mm}
\section{Methods}
\vspace{-1mm}
Denote a DWI volume as $\boldsymbol{I}\in \mathbb{R}^{T\times H\times W \times C}$, where $T$ is the number of DWI slices, $H$ and $W$ are the spatial dimensions (in pixels) of each 2D slice, respectively, and $C$ is the number of DWI channels. Our goal is to predict the segmentation map $\boldsymbol{O}\in \mathbb{R}^{T\times H\times W}$ of stroke lesions. The spatial resolution within each slice is 1 mm between adjacent pixels while the inter-slice resolution is 6 mm between slices. We can observe that the inter-slice resolution of DWIs is much lower than the intra-slice resolution, which leads to the high discontinuity between adjacent slices---the main characteristic of 2.5D data like DWIs. As discussed in Sec.~\ref{sec:intro}, both 3D and 2D segmentation models are not ideal for DWIs, because common 3D models are likely to overfit irrelevant features in neighboring slices, while 2D models completely disregard 3D contextual information. This motivates us to propose the \texttt{LambdaUNet}, a 2.5D segmentation model specifically designed for DWIs. Below, we will first provide an overview of \texttt{LambdaUNet} and then elaborate on how its \texttt{Lambda+} layers effectively capture 2.5D contextual features.  

\vspace{2mm}
\noindent\textbf{LambdaUNet.} The main structure of our \texttt{LambdaUNet} follows the UNet~\cite{ronneberger2015u} for its strong ability to preserve both high-level semantic features and low-level details. 
The key difference of \texttt{LambdaUNet} from the original UNet is that we replace convolutional layers in the UNet encoder with our proposed \texttt{Lambda+} layers (detailed in Sec.~\ref{sec:LambdaLayer}), which can extract both dense intra-slice features and sparse inter-slice features effectively. Since all layers except \texttt{Lambda+} layers in \texttt{LambdaUNet} are identical with those in UNet, they require 2D features as input; we address this by merging the slice dimension $T$ with the batch dimension to reshape 3D features into 2D features for non-\texttt{Lambda+} layers, while \texttt{Lambda+} layers undo this reshaping to recover the slice dimension and regenerate a 3D input that is used to extract both intra- and inter-slice features. The final output of \texttt{LambdaUNet} is the lesion segmentation mask $\boldsymbol{O}\in \mathbb{R}^{T\times H\times W}$. The Binary Cross-Entropy (BCE) loss is used to train \texttt{LambdaUNet} for the pixel-wise binary classification task.

\vspace{-4mm}
\subsection{Lambda+ Layers}
\label{sec:LambdaLayer}
\vspace{-1mm}
\texttt{Lambda+} layers are an enhanced version of Lambda layers~\cite{bello2021lambdanetworks}, which transform context around a pixel into linear functions, called \emph{lambdas}, and mimic the attention operation by applying lambdas to the pixel to produce features. Different from attention, the lambdas can encode positional information as we will elaborate later, which affords them a stronger ability to model spatial relations. \texttt{Lambda+} layers extend Lambda layers, which are designed for 2D data, by adding inter-slice lambdas with a restricted context region to effectively extract features from 2.5D data such as DWIs.

\begin{figure}[t]
\includegraphics[width=\textwidth]{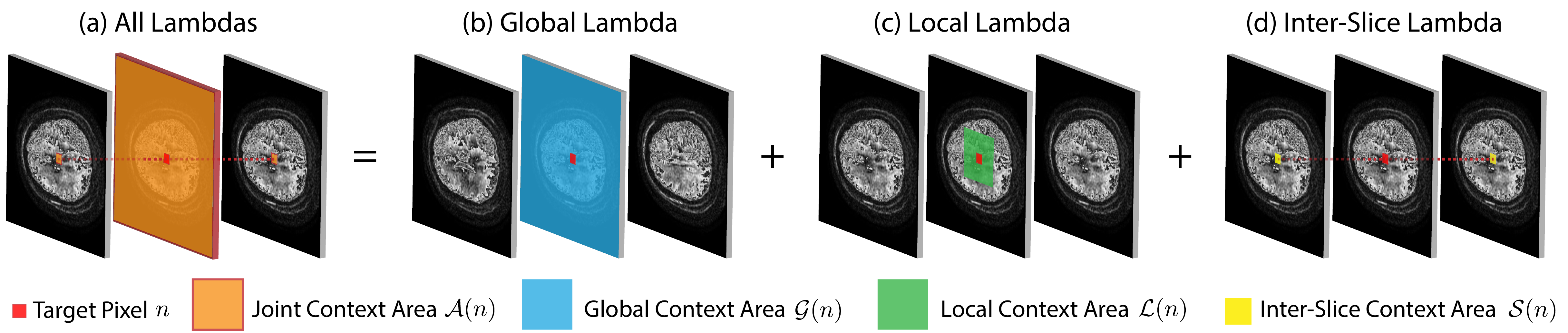}
\vspace{-6mm}
\caption{Context areas of the global lambda, local lambda, and inter-slice lambda.} \label{fig:lambdalayer}
\vspace{-4mm}
\end{figure}

The input to a \texttt{Lambda+} layer is a 3D feature map $\boldsymbol{X}\in\mathbb{R}^{|n| \times |c|}$, where $|c|$ is the number of channels and $n$ is the linearized pixel index into both spatial (height $H$ and width $W$) and slice ($T$) dimensions of the feature map, {\it i.e.}, $n$ iterates over all pixels $\mathcal{P}$ inside the 3D volume, and $|n|$ equals the total number of pixels $|\mathcal{P}|$. Besides input $\boldsymbol{X}$, we also have context $\boldsymbol{C} \in\mathbb{R}^{|m| \times |c|}$ where $\boldsymbol{C} = \boldsymbol{X}$ (same as self-attention) and $m$ also iterates over all pixels $\mathcal{P}$ in the 3D volume. Importantly, when extracting features for each pixel $n$, we restrict the region of context pixels $m$ to a 2.5D area $\mathcal{A}(n) \subset \mathcal{P}$. As shown in Fig.~\ref{fig:lambdalayer}\,(a), the 2.5D context area consists of the entire slice where pixel $n$ is in, as well as pixels with the same 2D location in adjacent $\mathcal{T}$ slices where $\mathcal{T}$ is the inter-slice kernel size.

Similar to attention, \texttt{Lambda+} layer computes queries $\boldsymbol{Q} = \boldsymbol{XW}_Q \in \mathbb{R}^{|n| \times |k|} $, keys $\boldsymbol{K} = \boldsymbol{CW}_K \in \mathbb{R}^{|m| \times |k| \times |u|} $, and values $\boldsymbol{V} = \boldsymbol{CW}_V  \in \mathbb{R}^{|m| \times |v| \times |u|} $, where $\boldsymbol{W}_Q\in\mathbb{R}^{|c|\times|k|}$, $\boldsymbol{W}_K\in\mathbb{R}^{|c|\times|k|\times|u|}$ and  $\boldsymbol{W}_V\in\mathbb{R}^{|c|\times|v|\times|u|}$ are learnable projection matrices, $|k|$ and $|v|$ are the dimensions of queries (keys) and values, and $|u|$ is an additional dimension to increase model capacity. We normalize the keys across pixels using softmax: $\bar{\boldsymbol{K}} = \text{softmax}(\boldsymbol{K})$. We denote $\boldsymbol{q}_n \in \mathbb{R}^{|k|}$ as the $n$-th query in $\boldsymbol{Q}$ for a pixel $n$. We also denote $\bar{\boldsymbol{K}}_m \in \mathbb{R}^{|k|\times|u|}$ and $\boldsymbol{V}_m \in \mathbb{R}^{|v|\times|u|}$ as the $m$-th key and value in $\boldsymbol{K}$ and $\boldsymbol{V}$ for a context pixel $m$.

For a target pixel $n \in \mathcal{P}$ inside a slice $t$, a \texttt{lambda+} layer computes three types of lambdas (linear functions) as illustrated in Fig.~\ref{fig:lambdalayer}: (1) a \textbf{global lambda} that encodes global context within slice $t$, (2) a \textbf{local lambda} that summarizes the local context around pixel $n$ in slice $t$, and (3) an \textbf{inter-slice lambda} that captures inter-slice features from adjacent slices.

\vspace{2mm}
\noindent\textbf{Global Lambda.} As shown in Fig.~\ref{fig:lambdalayer}\,(b), the global lambda aims to encode the global context within slice $t$ where the target pixel $n$ is in, so the context area $\mathcal{G}(n)$ of the global lambda includes all pixels within slice $t$. For each context pixel $m \in \mathcal{G}(n)$, its contribution to the global lambda is computed as:
\begin{equation}
    \boldsymbol{\mu}_{m}^{\texttt{G}}=\Bar{\boldsymbol{K}}_{m} \boldsymbol{V}_{m}^{T}\,, \quad m \in \mathcal{G}(n)\,.
\vspace{-1mm}
\end{equation}
The global lambda $\boldsymbol{\lambda}_n^{\texttt{G}}$ is the sum of the contributions from each pixel $m \in \mathcal{G}(n)$:
\vspace{-1mm}
\begin{equation}
    \boldsymbol{\lambda}_n^{\texttt{G}}=\sum_{m \in \mathcal{G}(n)} \boldsymbol{\mu}_{m}^{\texttt{G}}=\sum_{m \in \mathcal{G}(n)} \Bar{\boldsymbol{K}}_{m} \boldsymbol{V}_{m}^{T}\in \mathbb{R}^{|k| \times|v|}\,.
\vspace{-1mm}
\end{equation}
Note that $\boldsymbol{\lambda}_n^{\texttt{G}}$ is invariant for all $n$ within the same slice as $\mathcal{G}(n)$ is the same.

\vspace{2mm}
\noindent\textbf{Local Lambda.} The local lambda encodes the context of a local $R \times R$ area $\mathcal{L}(n)$ centered around the target pixel $n$ in slice $t$ (see Fig.~\ref{fig:lambdalayer}\,(c)). Compared with the global lambda, besides the difference in context areas, the local lambda uses \emph{learnable} relative-position-dependent weights $\boldsymbol{E}_{nm} \in \mathbb{R}^{|k|\times|u|}$ to encode the position-aware contribution of a context pixel $m$ to the local lambda:
\begin{equation}
    \boldsymbol{\mu}_{nm}^{\texttt{L}}=\boldsymbol{E}_{nm} \boldsymbol{V}_{m}^{T}\,, \quad m \in \mathcal{L}(n)\,.
\end{equation}
Note that the weights $\boldsymbol{E}_{nm}$ are shared for any pairs of pixels $(n, m)$ with the same relative position between $n$ and $m$.
The local lambda $\boldsymbol{\lambda}^{\texttt{L}}$ is obtained by:
\begin{equation}
    \boldsymbol{\lambda}_{n}^{\texttt{L}}=\sum_{m \in \mathcal{L}(n)} \boldsymbol{\mu}_{nm}^{\texttt{L}}=\sum_{m \in \mathcal{L}(n)} \boldsymbol{E}_{nm} \boldsymbol{V}_{m}^{T}\in \mathbb{R}^{|k| \times|v|}\,.
\vspace{-1mm}
\end{equation}

\vspace{2mm}
\noindent\textbf{Inter-Slice Lambda.} The inter-slice lambda defines a context area $\mathcal{S}(n)$ including pixels in adjacent slices sharing the same 2D location with the target pixel $n$, as shown in Fig.~\ref{fig:lambdalayer}\,(d). As discussed before, we use this restricted context area for extracting inter-slice features due to the high discontinuity between slices for 2.5D data like DWIs. Although one context pixel per adjacent slice seems very restrictive, one pixel of a feature map at coarse (downsampled) 2D scales in \texttt{LambdaUNet} corresponds to a large area in the original scale. Furthermore, \texttt{LambdaUNet} employs multiple \texttt{Lambda+} layers, so information from other pixels in adjacent slices can first propagate to pixels in $\mathcal{S}(n)$ and then to the target pixel $n$. Thus, our design of the restricted context area makes the network focus on the most informative pixels inside $\mathcal{S}(n)$ and suppress less relevant pixels, while still allowing pixels outside the area to indirectly contribute to the feature through multiple \texttt{Lambda+} layers.

Similar to the local lambda, the inter-slice lambda $\boldsymbol{\lambda}_{n}^{\texttt{S}}$ uses learnable weights $\boldsymbol{F}_{nm} \in \mathbb{R}^{|k|\times|u|}$ to encode position-aware contribution of context pixels:
\begin{equation}
    \boldsymbol{\mu}_{n m}^{\texttt{S}}=\boldsymbol{F}_{nm} \boldsymbol{V}_{m}^{T}\,,\quad m \in \mathcal{S}(n)\,,
\end{equation}
\begin{equation}
    \boldsymbol{\lambda}_{n}^{\texttt{S}}=\sum_{m \in \mathcal{S}(n)} \boldsymbol{\mu}_{n m}^{\texttt{S}}=\sum_{m \in \mathcal{S}(n)} \boldsymbol{F}_{n m} \boldsymbol{V}_{m}^{T}\in \mathbb{R}^{|k| \times|v|}\,.
\end{equation}

\vspace{2mm}
\noindent\textbf{Applying Lambdas.} After computing the global lambda $\boldsymbol{\lambda}_n^{\texttt{G}}$, local lambda $\boldsymbol{\lambda}_n^{\texttt{L}}$, and inter-slice lambda $\boldsymbol{\lambda}_n^{\texttt{S}}$, we are ready to apply them to the query $\boldsymbol{q}_n$ of the target pixel $n$. The output feature $\boldsymbol{y}_n$ for the target pixel $n$ is:
\begin{equation}
    \boldsymbol{y}_{n}=\boldsymbol{q}_{n}^T\left(\boldsymbol{\lambda}_n^{\texttt{G}}+\boldsymbol{\lambda}_{n}^{\texttt{L}}+\boldsymbol{\lambda}_{n}^{\texttt{S}}\right)\in \mathbb{R}^{|v|}\,.
\end{equation}
The final output of \texttt{Lambda+} layer is a 3D feature map $\boldsymbol{Y}\in\mathbb{R}^{|n| \times |v|}$ formed by the output features $\boldsymbol{y}_{n}$ of all pixels $n \in \mathcal{P}$. Although the above procedure for computing lambdas is for a single pixel $n$, we can easily parallelize the computation for all pixels using standard convolution operations, which makes \texttt{Lambda+} layers computationally-efficient. We refer readers to the pseudocode in Appendix~\ref{Pseudocode} for detailed implementation.

\vspace{-4mm}
\section{Experiments}
\label{Sec: experiments}
\vspace{-1mm}
The primary focus of our experiments is to answer the following questions: (1) Does \texttt{LambdaUNet} predict lesion segmentation maps more accurately than baselines? (2) Is our 2.5D \texttt{Lambda+} layer more effective than the 2D or 3D Lambda layer? (3) Based on qualitative results, does \texttt{LambdaUNet} has clinical significance?

\vspace{2mm}
\noindent\textbf{Dataset.}
The clinical data we use to evaluate our model is provided by an urban academic hospital. We sampled 99 acute ischemic stroke cases with large ($n=42$) and small ($n=57$) infarct size. The data has an equal distribution of samples from stroke with the left or right middle cerebral artery (MCA), posterior cerebral artery (PCA), and anterior cerebral artery (ACA) origins. The cases contain a mix of 1.5T and 3.0T scans. Certain cases even have a mix of MCA and ACA.  The ischemic infarcts are manually segmented by three experts based on diffusion-weighted imaging (DWI) (b=1000 s/mm2) and the calculated exponential apparent diffusion map (eADC) using MRICro v1.4. We use the eADC and DWI images from ischemic stroke patients to form the two channels of input DWIs $\boldsymbol{I}$.
We use 67 of the 99 fully-labeled cases for training and the remaining 32 fully-labeled cases for validation and testing. More specifically, we split the 32 cases into three folds of roughly the same size. Two of the three folds are used for validation and one remaining fold is used for testing. Each of the three folds is used for testing once, and the average result is reported as the final testing result. The 32 cases used for testing were carefully chosen to make sure the stroke size, location, and type are nicely balanced in the testing set.

\vspace{2mm}
\noindent\textbf{Implementation Details.}
Our implementation is using the PyTorch~\cite{paszke2019pytorch} and the Lightning~\cite{falcon2019pytorch} frameworks. All experiments are conducted using four NVIDIA Quadro RTX 6000 GPUs with 24 GB memory. For \texttt{Lambda+} layers, both the inter-slice kernel size $\mathcal{T}$ and the local kernel size $R$ are set to 3. We train the model for 100 epochs using the RMSprop optimizer; an initial learning rate of \texttt{1e-4} is used for 20 epochs and then the learning rate is linearly reduced to 0. We randomly select 12 DWI sequence segments of 8 slices to form a mini-batch during training. The whole training process takes about 4 hours to finish. The training converges after 40 epochs. For testing, we select the model that gives the highest dice score for validation data.

\vspace{2mm}
\noindent\textbf{Baselines and Metrics.}
We compare our method against well-known and recent 2D segmentation methods, U-Net~\cite{ronneberger2015u}, AttnUNet~\cite{oktay2018attention}, and TransUNet~\cite{chen2021transunet}, as well as one 3D segmentation method: 3D UNet~\cite{cciccek20163d}. All the baseline methods are reproduced based on their open-sourced code with careful hyperparameter tuning. 
Besides, we also report the results of two variants of \texttt{LambdaUNet} to further evaluate the effectiveness of the proposed 2.5D lambda+ layer.
We use four common evaluation metrics---dice score coefficient (DSC), recall, precision, and $F_1$ score---for stroke lesion segmentation to provide quantitative comparisons.

\vspace{-2mm}
\subsection{Results}
In the first group of Table~\ref{tab:baseline}, we show the quantitative results of all baselines on our stroke lesion dataset. One can observe that the proposed \texttt{LambdaUNet} has significant improvements over baselines, {\it e.g.}, performance gains range from 3.06\% to 8.31\% for average DSC. The improvement suggests that our \texttt{Lambda+} layers are more suitable for feature extraction of 2.5D DWI data.
In the second group, we compare \texttt{LambdaUNet} with its 2D and 3D variants. \texttt{LambdaUNet2D} directly removes the inter-slice lambda from the \texttt{LambdaUNet} while \texttt{LambdaUNet3D} uses a 3D local context area $\mathcal{L}(n)$ instead of the inter-slice lambda. As indicated in Table~\ref{tab:baseline}, both variants perform worse than \texttt{LambdaUNet} in terms of DSC and the $F_1$ score. This demonstrates the effectiveness of the 2.5D design of the proposed \texttt{Lambda+} layers. Although \texttt{LambdaUNet} does not achieve the highest precision or recall over the baselines and variants, it can maintain a good balance between recall and precision, which sometimes cancel each other out ({\it e.g.}, AttnUNet and 3D UNet). This is further confirmed by the superior $F_1$ score of \texttt{LambdaUNet}.

\begin{table}[b]
\vspace{-3mm}
\caption{Segmentation performance comparison between different models.}\label{tab:baseline}
\vspace{-5mm}
\begin{center}
\begin{tabular}{@{\hskip 1mm}lccccccc@{\hskip 1mm}}
\toprule
Method & 2D/3D && DSC & &Recall/Precision && $F_1$ Score\\ 
\midrule
UNet~\cite{ronneberger2015u} & 2D && 82.15 && 80.28/86.29 && 81.61 \\
AttnUNet~\cite{oktay2018attention} & 2D && 81.83  && 77.45/86.74 && 80.82 \\
TransUNet~\cite{chen2021transunet} & 2D && 83.45 && 83.24/87.15 && 84.48 \\
3D UNet~\cite{cciccek20163d} & 3D && 78.20 && \textbf{83.54}/78.39 && 78.21 \\
\midrule
(Ours) LambdaUNet-2D & 2D && 84.03 &&  82.27/87.10  && 84.19 \\
(Ours) LambdaUNet-3D & 3D && 84.76  && 79.92/\textbf{89.86} && 84.09\\
\textbf{(Ours) LambdaUNet} & 2.5D && \textbf{86.51} &&  81.76/89.39 && \textbf{84.84}\\
\bottomrule
\end{tabular}
\end{center}
\end{table}

\begin{figure}[t!]
\vspace{-2mm}
\includegraphics[width=\textwidth]{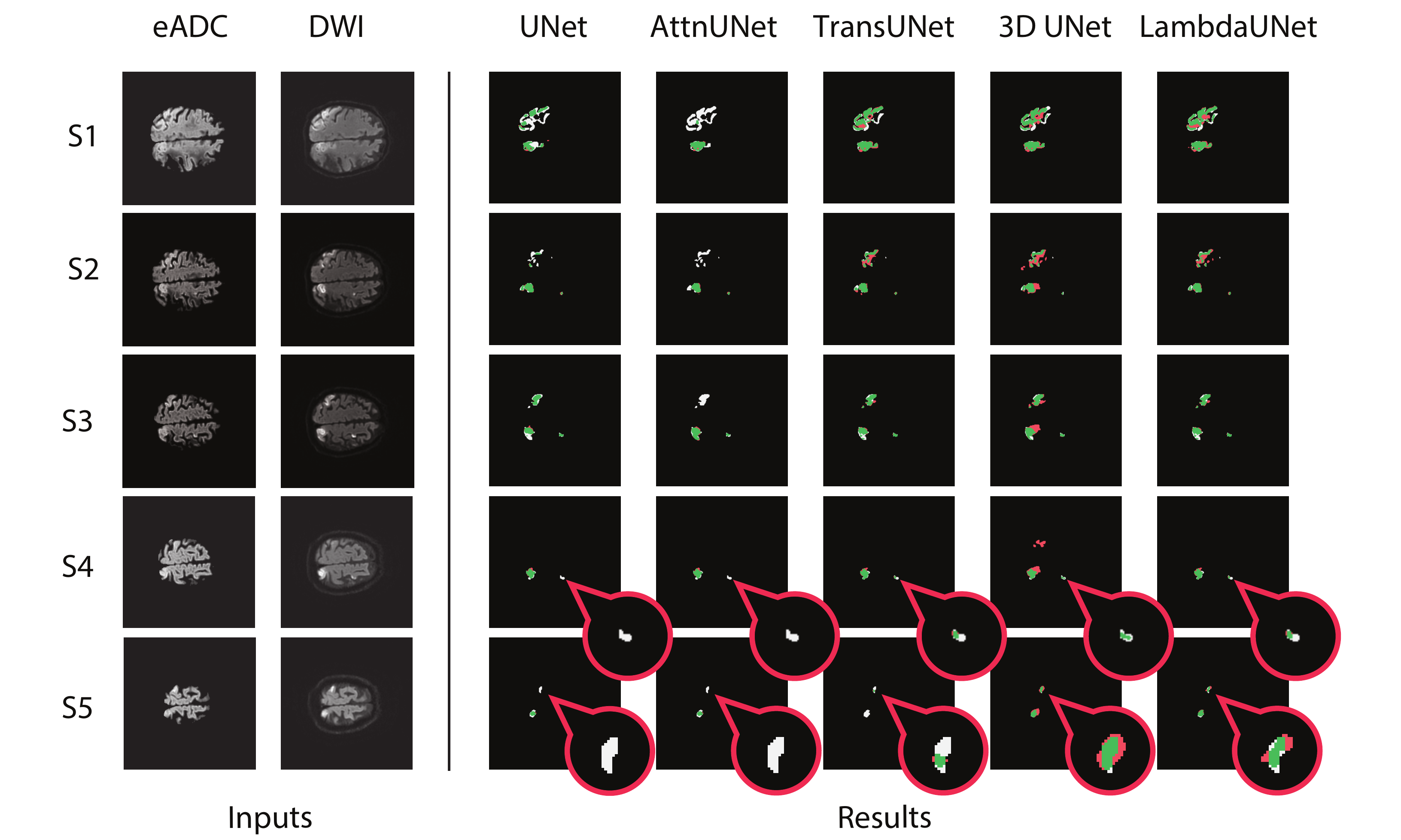}
\vspace{-6mm}
\caption{Qualitative results on five consecutive slices of one ischemic stroke clinical case. Green indicates the correct predictions. White areas are false negative while red areas are false positive. Red circles show a close-up view of the lesion areas.} \label{fig:results}
\vspace{-3mm}
\end{figure}

Fig.~\ref{fig:results} visualizes the predicted segmentation masks on five consecutive slices for one stroke case. We can see that the masks produced by our \texttt{LambdaUNet} (last column) are the closest to the ground truth than the baselines. For instance, in slice 3 (S3), the baselines either miss some details (UNet, AttnUNet, TransUNet) indicated by the white areas or generate some false positive predictions (3D UNet) denoted by the red areas, while our \texttt{LambdaUNet} captures the irregular shape of lesions well. S4 and S5 also show that \texttt{LambdaUNet} performs the best on difficult small lesions. More results are provided in Appendix~\ref{additional results}. 

\vspace{-2mm}
\subsection{Discussion}
Our \texttt{LambdaUNet} not only shows advantages on both quantitative and qualitative measurements, the way it extracts features is more like clinicians. As clinicians consider all adjacent slices but only focus on the most informative areas, our \texttt{Lambda+} layers capture intra- and inter-slice features and automatically suppress irrelevant 3D interference. Lesion areas of acute stroke are an important end-point for clinical trials, as proper treatment relies on measuring the infarction core volume and estimating salvageable tissue. Therefore, an accurate and reproducible DWI-suited segmentation model like \texttt{LambdaUNet} will be of high interest in clinical practice. 

\vspace{-2mm}
\section{Conclusion}
\vspace{-1mm}
We defined DWIs as 2.5D data for its dense intra-slice resolution and sparse inter-slice resolution. Based on the 2.5D characteristics, we proposed a segmentation network \texttt{LambdaUNet}, which includes a new 2.5D feature extractor, termed \texttt{Lambda+} layers. \texttt{Lambda+} layers effectively capture features in 2.5D data by using dense intra-slice and sparse inter-slice context areas. This design allows the network to focus on informative features while suppressing less relevant features to reduce noise and improve generalization. Experiments on the clinical stroke dataset verify that our \texttt{LambdaUNet} outperforms state-of-the-art segmentation methods and shows strong potential in clinical practice.

\bibliographystyle{splncs04}
\bibliography{refs}

\newpage
\appendix
\section{Pseudocode for Lambda+ Layers}
\vspace{-3mm}
\label{Pseudocode}
\begin{lstlisting}[language=Python, caption=Pseudocode for Lambda+ Layers]
class LambdaPlusLayer:
    def __init__(self):
        # c: input channel, k: queries (keys) dimensions, v: value dimensions
        # u: intra-depth, R: intra-slice kernel size, T: inter-slice kernel size
        to_q = Conv2dLayer(c, k, kernel=(1, 1)) # W_Q are weights in to_q
        to_k = Conv2dLayer(c, k * u, kernel=(1, 1)) # W_K are weights in to_k
        to_v = Conv2dLayer(c, k * u, kernel=(1, 1)) # W_V are weights in to_v
        local_lambda = Conv3dLayer(u, k, kernel=(1, R, R)) # E_{nm} are Conv weights
        inter_slice_lambda = Conv2dLayer(u, k, kernel=(1, T)) # F_{nm} are Conv weights
        
    def forward(self, x):
        # b: batch size, t: number of slices, hh: height, ww: width
        # x: input of Lambda+ Layer with size of (b * t) * hh * ww * c
        # B: b * t, merging slice dimension t with batch dimension b
        p = hh * ww # number of 2D pixels
        q, k, v = to_q(x), to_k(x), to_v(x)
        q = rearrange(q, 'B k hh ww -> B k (hh ww)')
        k = rearrange(k, 'B (u k) hh ww -> B u k (hh ww)')
        v = rearrange(v, 'B (u v) hh ww -> B u v (hh ww)')
        k = softmax(k, dim=-1)
        # Global Lambda
        lambda_g = einsum('B u k m, B u v m -> B k v', k, v)
        lambda_g = lambda_g.unsqueeze(-1).repeat(p, dim=-1) # shape: (B k v p)
        # Local Lambda
        v_l = reshape(v, 'B u v (hh ww) -> B u v hh ww')
        lambda_l = local_lambda(v_l)
        lambda_l = reshape(lambda_l, 'B k v hh ww -> B k v (hh ww)')
        # Inter-Slice Lambda
        v_s = rearrange(v, '(b t) u v p -> (b p) u v t')
        lambda_s = inter_slice_lambda(v_s)
        lambda_s = reshape(lambda_s, '(b p) k v t -> (b t) k v p')
        lambda = lambda_g + lambda_l + lambda_s
        Y = einsum('B k p, B k v p -> B v p', q, lambda)
        return Y
        
\end{lstlisting}
\vspace{3mm}

\section{Ablation Study of Inter-Slice Kernel Size}
\vspace{-4mm}
\begin{table}[h!]
\caption{Effect of inter-slice kernel size $\mathcal{T}$}\label{tab:baseline}
\vspace{-3mm}
\begin{center}
\begin{tabular}{@{\hskip 1mm}lccccccc@{\hskip 1mm}}
\toprule
 $\mathcal{T}$  && DSC & &Recall/Precision && $F_1$ Score\\ 
\midrule
3 && \textbf{86.51} &&  81.76/\textbf{89.39} && 84.84 \\
5 && 85.75  && \textbf{82.73}/88.80 && \textbf{85.22}\\
7 && 86.01 &&  81.68/88.70 && 84.60\\
\bottomrule
\end{tabular}
\end{center}
\end{table}

\clearpage
\section{Additional Qualitative Results}
\label{additional results}
\vspace{-6mm}
\begin{figure}[h!]
\includegraphics[width=\textwidth]{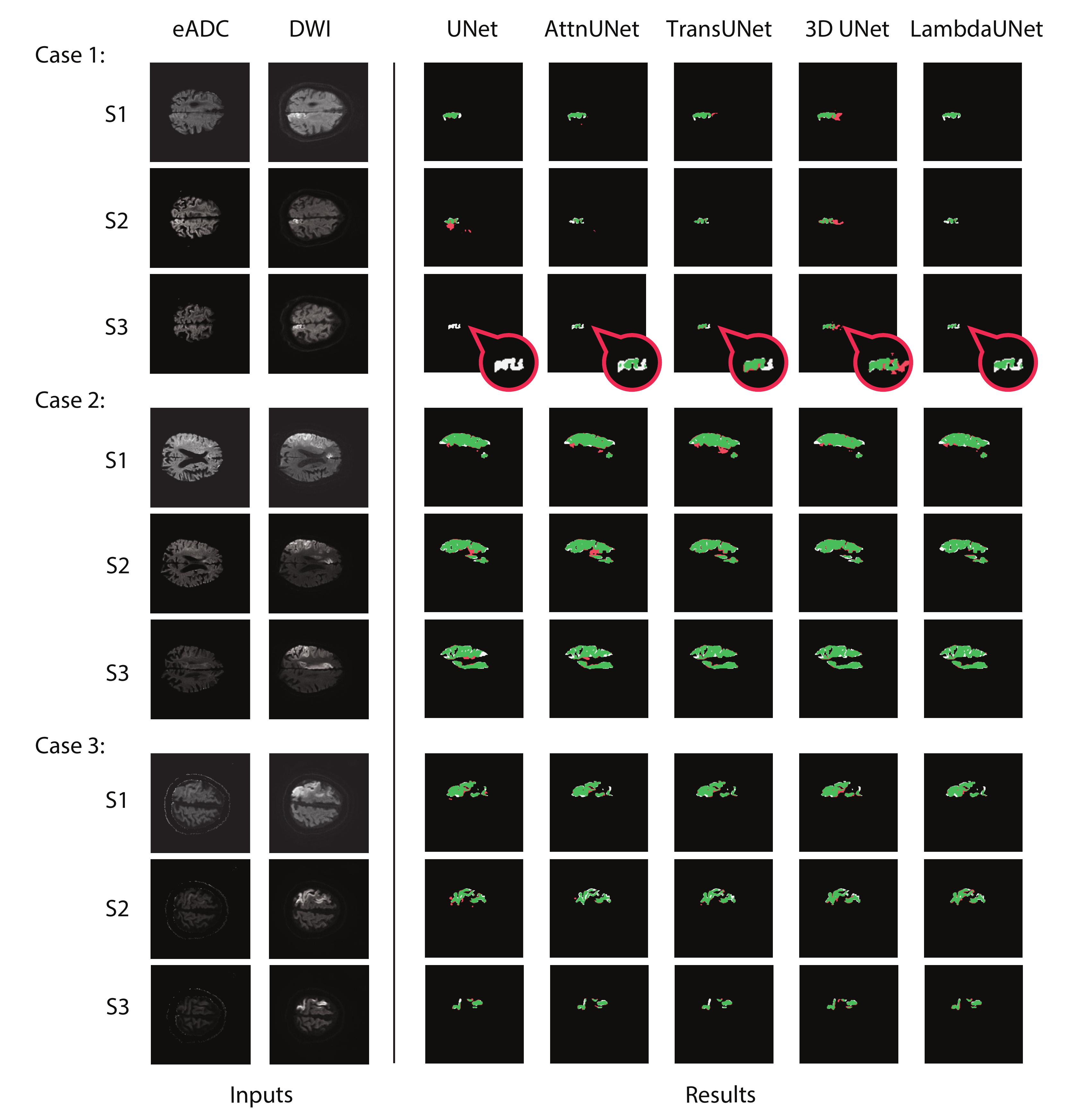}
\vspace{-7mm}
\caption{Qualitative results on three ischemic stroke clinical cases with consecutive slices. Green indicates the correct predictions. White areas are false negative while red areas are false positive. Red circles show a close-up view of the lesion areas.} \label{fig:supp}
\vspace{-3mm}
\end{figure}

\end{document}